\documentclass[letterpaper,english,aps,prx,superscriptaddress,twocolumn,amsfonts, amssymb]{revtex4}
\usepackage[T1]{fontenc}
\pdfoutput=1
\usepackage{textcomp}
\usepackage{mathrsfs}
\usepackage{amsmath}
\usepackage{amssymb}
\usepackage{graphicx}
\usepackage{esint}

\makeatletter
\usepackage[english]{babel}

\usepackage[bookmarks=true,colorlinks,linkcolor=blue,urlcolor=blue,citecolor=blue]{hyperref}

\newcommand{\be}{\begin{equation}}
\newcommand{\ee}{\end{equation}}
\newcommand{\bea}{\begin{eqnarray}}
\newcommand{\eea}{\end{eqnarray}}

\usepackage{epsfig,graphicx,psfrag,amsmath,amssymb,float}
\usepackage[caption=false]{subfig}

\@ifundefined{showcaptionsetup}{}{%
 \PassOptionsToPackage{caption=false}{subfig}}
\usepackage{subfig}
\makeatother

\usepackage{babel}

\begin{document}
\title{Orbital Magnetic Field Effects in Mott Insulators with Strong Spin-Orbit Coupling}
\author{Willian M. H. Natori}
\affiliation{Blackett Laboratory, Imperial College London, London SW7 2AZ, United
Kingdom}
\author{Roderich Moessner}
\affiliation{Max-Planck-Institut fur Physik komplexer Systeme, N\"othnitzer Stra{\ss}e 38, 01187 Dresden, Germany}
\author{Johannes Knolle}
\affiliation{Blackett Laboratory, Imperial College London, London SW7 2AZ, United
Kingdom}
\begin{abstract}
We study the effect of a magnetic field on the low energy description of Mott insulators with  strong spin-orbit (SO) coupling. In contrast to the standard case of the Hubbard model without SO coupling, we show that Peierls phases can modulate the magnetic exchange  at leading order in the interaction.  
Our mechanism crucially depends on the existence of distinct
exchange paths between neighboring magnetic ions enclosing a 
well-defined area. Thus it will generically be present in any
solid state realisation of the Kitaev model and its extensions.
We explicitly calculate the variation of the exchange constants of the so-called
$JK\Gamma$ model as a function of the magnetic flux.  We
discuss  experimental implications of our findings for various settings of candidate Kitaev spin liquids. 
\end{abstract}

\maketitle

Effective low energy descriptions have been crucial for advancing our understanding of correlated quantum phenomena in condensed matter physics as the complexity of microscopic Hamitonians is dramatically reduced. The canonical example is the derivation of the Heisenberg model~\cite{Anderson1959} with exchange constant $J\sim t^2/U$ from the half filled single band Hubbard model with kinetic energy $t$ and on-site repulsion $U$ in the Mott insulator (MI) limit~\cite{Takahashi1977}. Thouless and later Takahashi realised that charge fluctuations beyond the leading order in the interaction give rise to new higher order ring exchange terms~\cite{Thouless1965,Takahashi1977}. These have been extensively discussed in the early context of cuprate high temperature superconductivity~\cite{MacDonald1988} where they capture the increased role of charge-, hence, effective quantum-fluctuations when approaching the Mott transition. Moreover, higher order spin exchange terms  may  stabilise sought after quantum spin liquid (QSL) phases~\cite{Misguich1999,LiMing2000,Delannoy2005,Yang2010}.

An interesting observation for Hubbard models on non-bipartite lattices was that the application of a magnetic field changes the effective low energy model because charge fluctuations enclosing real space areas lead to higher order spin interactions sensitive to the enclosed magnetic flux~\cite{Sen1995}. It was shown that the Peierls phases of the orbital magnetic field induce a three-spin scalar chirality term~\cite{Sen1995} at order $t^3/U^2$ which is odd under time-reversal symmetry (TRS) and which may again stabilise a QSL phase~\cite{Motrunich2006}.  In recent years the long search for QSLs~\cite{Balents2010,Knolle2019} has  concentrated on MIs in the strong spin-orbit (SO) coupling limit~\cite{Rau2016,Hermanns2018,Winter2017} following the proposal~\citep{Jackeli2009} that they might be described by the  Kitaev honeycomb model (KHM) at low energies~\citep{Kitaev2006}.
 
Here, we investigate how the low energy description of MI with strong SO coupling is modified in  presence of a magnetic field. We discover that in contrast to the Hubbard model, the exchange constants are modified to leading order in the interactions; and that TRS odd terms such as a diamagnetic Zeeman term appear also for the bipartite honeycomb lattice. 
For most Kitaev transition metal candidates, such as $\alpha$-RuCl$_{3}$, the modulation turns out to be small -- at most relevant in the pulsed magnetic field regime to study the polarized phase --  because the small area between magnetic ions leads only to a small value of enclosed flux. However, our basic mechanism will be relevant for the recent Kitaev material proposal in metal-organic frameworks (MOFs) \citep{YamadaPRL2017,YamadaPRB2017}, where the presence of organic ligands 
leads to larger distances between the magnetic ions, and hence considerably increased fluxes enclosed by the exchange pathways.

\begin{figure}
\begin{centering}
\subfloat{\includegraphics[width=0.4\columnwidth]{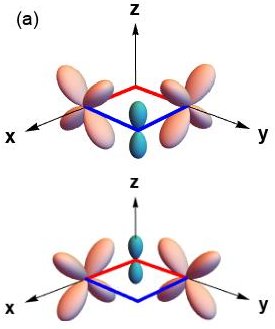}}\quad\subfloat{\includegraphics[width=0.4\columnwidth]{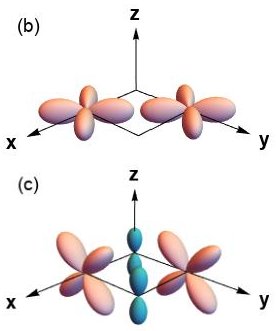}}
\par\end{centering}
\caption{\label{fig:T} (Color online) Active orbitals on the $xy$-plane for
the tight-binding Hamiltonians (a) $\hat{T}_{2}$, (b) $\hat{T}_{3}$,
and (c) $\hat{T}_{1}$. The path assigned in blue (red) gives rise
to the phase $\phi_{1}$ ($\phi_{2}$) discussed in the main text.}
\end{figure}

\begin{figure}
\begin{centering}
\subfloat[\label{fig:edge_sharing}]{\includegraphics[width=0.35\columnwidth]{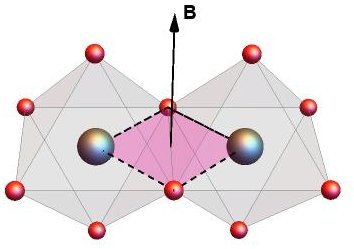}

}\subfloat[\label{fig:ligand}]{\includegraphics[width=0.6\columnwidth]{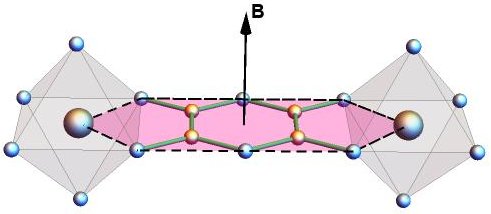}

}
\par\end{centering}
\caption{\label{fig:Areas} (Color online) Neighboring octahedra containing
the magnetic ions (a) in the originally proposed Kitaev materials and 
(b) in metal-organic frameworks. The magnetic flux through the $\mathcal{A}$
areas highlighted in magenta gives rise to the Peierls phase in Eq.
(\ref{eq:phi}).}
\end{figure}

Our paper is organised as follows. First, we provide the necessary background on the physics of magnetic exchange in 
our class of SO-coupled magnets, specifically introducing the so-called $JK\Gamma$ model. We then
include the effect of the magnetic field
by applying the Peierls substitution on the hopping integrals of the
transfer processes indicated in Fig.~\ref{fig:T}.  
Projecting to the low energy manifold we obtain a modified
model with  couplings dependent on the magnetic flux
over areas $\mathcal{A}$ depicted in Fig.~\ref{fig:edge_sharing}. 
Next, we discuss the resulting modulation of the exchanges. Finally, 
we assess its relevance for tuning currently available Kitaev materials into 
QSL phases and for cold atom proposals of the KHM.

Our starting point is the 
seminal paper Jackeli and Khaliullin, who argued that the KHM emerges at low energies in transition metal compounds
with 4$d^{5}$ or 5$d^{5}$ magnetic ions (M) with strong SO coupling~\citep{Jackeli2009}.
Each transition metal ion is surrounded by six anions (X) forming
a perfectly octahedral environment MX$_{6}$, such that the site is
described by a single hole occupying one $t_{2g}$ orbital. Ref. \citep{Jackeli2009}
dedicated special attention to the case of compounds with edge-sharing
MX$_{6}$ octahedra, since in this case the M ions can form honeycomb
lattices. Symmetry constraints between the orbital shapes and octahedral
arrangements motivated the study of superexchange processes driven
by the hole transfers occurring along the two distinct paths displayed
in figure \ref{fig:T}(a). Second-order perturbation theory is
then used to derive effective Kugel-Khomskii models \citep{Kugel1982,Khaliullin2005}
for each of the possible exchange paths, which are then projected
to a $j=1/2$ Hilbert space defined by the strong SO coupling.
The resulting Hamiltonians in each path are quantum compass models
\citep{Nussinov2015} that combine the KHM and Heisenberg interactions.
With time reversal symmetry the sum of all these compass models  cancels the Heisenberg
terms and retains only the Kitaev model. This we show is no longer the case in the presence of a magnetic field. 

Following Ref.~\citep{Jackeli2009} several compounds were synthesised, but they often display  long range magnetic order~\citep{Winter2017,Hermanns2018}.
In fact, the magnetism of these so-called \emph{Kitaev Materials}
is not described by the pure KHM as other hole-transfer processes lead to more complicated effective Hamiltonians such as
the Heisenberg-Kitaev (HK) model \citep{Chaloupka2010,Chaloupka2013}
and the $JK\Gamma$ model. Here, we concentrate on the derivation of the $JK\Gamma$ model as a minimal microscopic model of Kitaev materials combining Kitaev, Heisenberg
and symmetric off-diagonal exchanges \citep{Rau2014,Winter2016}.

Studies of the
$JK\Gamma$ model \citep{Rau2014,Winter2016} showed that a QSL 
would be stable only for materials that are very close to the idealised situation 
described in Ref.\citep{Jackeli2009}. To suppress the residual magnetism and induce a genuine QSL phase a promising route is to tune parameters of the $JK\Gamma$ model
via external fields such as pressure~\citep{Veiga2017,Yadav2018} or via
an applied  magnetic field $\mathbf{B}$. The latter route to a QSL is being
investigated theoretically~\cite{PhysRevB.97.241110,PhysRevB.98.014418,jiang2018field,zou2018field,patel2018magnetic,hickey2019emergence,jiang2019tuning,kaib2019kitaev} and experimentally in the candidate material $\alpha$-RuCl$_{3}$, which displays a zigzag order
in the absence of a magnetic field $\mathbf{B}$ \citep{PlumbPRB2014,Sears2015,Johnson2015}
but an apparently disordered state when subject to $B\sim14$T \citep{Sears2017,Baek2017,Hentrich2018,Banerjee2016,Banerjee2017,Banerjee2018}. A central question which we address  is whether the magnetic field induces additional changes beyond the commonly assumed simple Zeeman term.
As the central result we find that a magnetic field modulates the
exchange couplings of the $JK\Gamma$ model and induces a diamagnetic Zeeman term in Kitaev materials.

\emph{Model Derivation.} 
We begin with the study of the Peierls substitution
effects on the original Jackeli-Khaliullin mechanism. The underlying
multi-orbital Hubbard model for Kitaev materials is

\begin{equation}
H_{\text{Hub}}=\hat{V}+\hat{T}_{2},
\end{equation}
where $\hat{V}$ gives the onsite interactions among $t_{2g}$ holes
that depend upon the Coulomb repulsion $U$ and the Hund's coupling
$J_{H}$ \citep{Kugel1982,Khaliullin2005} while $\hat{T}_{2}$ is
the tight-binding Hamiltonian between the orbitals indicated in Fig.
\ref{fig:T}(a). $\hat{T}_{2}$ is a bond-dependent Hamiltonian, e.g.,
only the $\left|yz\right\rangle $ and $\left|zx\right\rangle $ orbitals
are active for bonds on $xy$-planes. The hopping integrals $t_{ij}$
in $\hat{T}_{2}$ are modified when the system is subject to a magnetic
field $\mathbf{B}$ \citep{Peierls1933}. Within the Peierls substitution, such modification
reads
\begin{equation}
t_{ij}\rightarrow t_{ij}\exp\left(\frac{ie}{\hbar c}\int_{\mathbf{r}_{i}}^{\mathbf{r}_{j}}d\mathbf{r}^{\prime}\cdot\mathbf{A}\left(\mathbf{r}^{\prime}\right)\right),\label{eq:Peierls}
\end{equation}
where $\mathbf{A}$ is the vector potential. In the remainder of the
paper, we restrict the discussion to a magnetic field applied in the
$\hat{\mathbf{z}}$ direction with $\mathbf{A}=\frac{B}{2}\left(-y,x,0\right)$.
We will also focus on minimal models involving two sites $i$ and
$j$ connected by a bond $\left\langle ij\right\rangle _{z}$ on the
$xy$ plane. The relevant terms of $\hat{T}_{2}$ for these two sites
are given by
\begin{align}
\hat{T}_{2}^{(1)} & =-t_{2}\underset{\sigma}{\sum}\left(d_{i,yz,\sigma}^{\dagger}d_{j,zx,\sigma}e^{i\phi_{1}}+d_{j,zx,\sigma}^{\dagger}d_{i,yz,\sigma}e^{-i\phi_{1}}\right),\nonumber \\
\hat{T}_{2}^{(2)} & =-t_{2}\underset{\sigma}{\sum}\left(d_{i,zx,\sigma}^{\dagger}d_{j,yz,\sigma}e^{i\phi_{2}}+d_{j,yz,\sigma}^{\dagger}d_{i,zx,\sigma}e^{-i\phi_{2}}\right),
\end{align}
with $t_{2}$ the hopping integral, the superscripts label the
two different superexchange paths, and $\phi_{i}$ are the Peierls
phases calculated after Eq. (\ref{eq:Peierls}). The effective model
obtained in the strong-coupling approach followed by the projection
on the $j=1/2$ manifold is the HK model:
\begin{align}
H_{\text{Pei},\left\langle ij\right\rangle _{z}}^{(1)} & =4\left[\nu_{1}-\nu_{2}-\left(\nu_{1}+2\nu_{2}\right)\cos\left(\phi_{1}-\phi_{2}\right)\right]s_{i}^{z}s_{j}^{z}\nonumber \\
 & \quad-2\nu_{1}\left[1-\cos\left(\phi_{1}-\phi_{2}\right)\right]\mathbf{s}_{i}\cdot\mathbf{s}_{j},\label{eq:HPeizKH}
\end{align}
in which the energy unity is set at $4t_{2}^{2}/\left(9U\right)$.
The constants $\nu_{1}=\left(3r_{1}+r_{2}+2r_{3}\right)/6$ and $\nu_{2}=\left(r_{1}-r_{2}\right)/4$
are dependent on the ratios $r_{1}=1/\left(1-3\eta\right)$, $r_{2}=1/\left(1-\eta\right)$
and $r_{3}=1/\left(1+2\eta\right)$, where $\eta=J_{H}/U$. The phase
difference, Eq. (\ref{eq:HPeizKH}), is  related to the magnetic
field through
\begin{equation}
\phi_{1}-\phi_{2}=\frac{e}{\hbar c}\oint_{\mathcal{C}}d\mathbf{r}^{\prime}\cdot\mathbf{A}\left(\mathbf{r}^{\prime}\right)=\frac{e}{\hbar c}\int_{\mathcal{A}}d\mathbf{a}\cdot\mathbf{B}\equiv2\pi\frac{\Phi}{\Phi_{0}}.\label{eq:phi}
\end{equation}
In Eq. (\ref{eq:phi}), $\Phi_{0}$ is the magnetic flux quantum and
$\Phi$ is the magnetic flux passing through the area $\mathcal{A}$
enclosed by the superexchange paths, as depicted in Fig. \ref{fig:Areas}.
The geometrical symmetry allows us to take $\phi_{2}=-\phi_{1}$ in
the chosen gauge, and we use this equality throughout this paper.
We point out that the magnetic flux piercing the $\mathcal{A}$ areas
related to bonds on $zx$ and $yz$ planes is zero, and the Heisenberg
interaction is absent in these cases. However, similar HK models will
be observed in other planes through the variation of the $\mathbf{B}$
direction.

We are now prepared to discuss the modulation of the $JK\Gamma$ model
after applying the Peierls substitution on tight-binding models that
include the additional hopping processes indicated in Fig. \ref{fig:T}(b)
and Fig. \ref{fig:T}(c) \citep{Rau2014,Winter2016,YamadaPRL2017}.
Fig. \ref{fig:T}(b) illustrates the active orbitals in the tight-binding
Hamiltonian $\hat{T}_{3}$. The microscopic details that distinguish
$\hat{T}_{3}$ for MOFs and transition metal compounds lead to different
ways of applying Peierls substitution. In MOFs, the hopping is mediated
by two different paths involving $\sigma$ molecular orbitals \citep{YamadaPRL2017}.
These paths follow closely the ones defined by the Jackeli-Khaliullin
mechanism, in such a way that the same phases $\phi_{1,2}$ can be
assigned to $\hat{T}_{3}$ under Peierls substitution. By contrast,
$\hat{T}_{3}$ in transition metal compounds is related to the direct
overlap between the orbitals without any alternative path, which makes
the phase defined in Eq. (\ref{eq:Peierls}) irrelevant. The two cases
can be written simultaneously by the Hamiltonian $\hat{T}_{3}=\hat{T}_{3}^{(1)}+\hat{T}_{3}^{(2)}$
given by

\begin{align}
\hat{T}_{3}^{(1)} & =-\frac{t_{3}}{2}\underset{\sigma}{\sum}\left(d_{i,xy,\sigma}^{\dagger}d_{j,xy,\sigma}e^{i\phi_{a}}+\text{h.c.}\right),\nonumber \\
\hat{T}_{3}^{(2)} & =-\frac{t_{3}}{2}\underset{\sigma}{\sum}\left(d_{i,xy,\sigma}^{\dagger}d_{j,xy,\sigma}e^{i\phi_{b}}+\text{h.c.}\right),
\end{align}
where $\phi_{a}=\phi_{1},\,\phi_{b}=\phi_{2}$ for MOFs and $\phi_{a}=\phi_{b}=0$
for transition metal compounds. Finally, the active orbitals encompassed
by $\hat{T}_{1}$ are illustrated in \ref{fig:T}(c) and occur when
there is a hybridization between the $p$-atomic or $\pi$-molecular
orbitals bridging the two sites. The equivalent of the $\mathbf{B}$-induced
phase in Eq. (\ref{eq:Peierls}) should be given by a path integral
that considers all classical paths connecting $i$ and $j$, which
must be zero by symmetry. The Hamiltonian $\hat{T}_{1}$ then reads
\begin{equation}
\hat{T}_{1}=-t_{1}\underset{\sigma}{\sum}\left(d_{i,yz,\sigma}^{\dagger}d_{j,yz,\sigma}+d_{i,zx,\sigma}^{\dagger}d_{j,zx,\sigma}+\text{h.c.}\right).
\end{equation}

Following the Jackeli-Khaliullin procedure for the tight-binding
Hamiltonian $\hat{T}=\hat{T}_{1}+\hat{T}_{2}+\hat{T}_{3}$ yields
\begin{align}
H_{\text{Pei},\langle ij\rangle_{z}}^{(2)} & =K\left(\phi_{1}\right)s_{i}^{z}s_{j}^{z}+\Gamma\left(\phi_{1}\right)\left(s_{i}^{x}s_{j}^{y}+s_{i}^{y}s_{j}^{x}\right)\nonumber \\
 & +J\left(\phi_{1}\right)\mathbf{s}_{i}\cdot\mathbf{s}_{j}+h\left(\phi_{1}\right)\left(s_{i}^{z}+s_{j}^{z}\right),\label{eq:JKG_model}
\end{align}
which is simply the $JK\Gamma$ model with modulated exchange constants
and an emergent diamagnetic term. After setting $x_{1}\equiv t_{1}/t_{2}$
and $x_{3}\equiv t_{3}/t_{2}$ we find the analytic expression for the modulated exchanges: \begin{subequations}
\begin{align}
J\left(\phi_{1}\right) & =4\left[\nu_{1}x_{1}^{2}-\nu_{1}\sin^{2}\phi_{1}+\frac{\nu_{1}-2\nu_{2}}{4}x_{3}^{2}\cos^{2}\phi_{a}\right.\nonumber \\
 & \quad\,\left.+\left(\nu_{1}+2\nu_{2}\right)x_{1}x_{3}\cos(\phi_{a})\right],\\
K\left(\phi_{1}\right) & =4\left\{ \nu_{2}\left[x_{1}^{2}+x_{3}^{2}\cos^{2}(\phi_{a})-2x_{1}x_{3}\cos(\phi_{a})\right]\right.\nonumber \\
 & \qquad\left.+\left[\nu_{1}-\nu_{2}-\cos(2\phi_{1})\left(\nu_{1}+2\nu_{2}\right)\right]\right\} ,\\
\Gamma\left(\phi_{1}\right) & =8\nu_{2}\cos\phi_{1}\left(x_{1}-x_{3}\cos\phi_{a}\right),\\
h\left(\phi_{1}\right) & =\left(r_{1}+r_{2}\right)\left(x_{1}-x_{3}\cos\phi_{a}\right)\sin\phi_{1}.
\end{align}
\label{JKG_constants_phi}\end{subequations}

Equations~\ref{JKG_constants_phi} present the main result of our work - a magnetic field modulation of the exchanges at leading order in the interaction. The emergent diamagnetic term linear in spin operators is somewhat surprising but a direct consequence of the interplay of SO coupling and the broken TRS via the Peierls phases. Note, a magnetic
field such as $h\left(\phi_{1}\right)$ has been previously shown to appear in quantum compass
model implementations of bosonic optical lattices \citep{Radi2012}.
There, it is in principle possible to tune the sign of the
emergent magnetic field but in contrast here, $h\left(\phi_{1}\right)$ in
Eq. (\ref{eq:JKG_model}) always opposes $\mathbf{B}$ and is directly
related to the diamagnetism in Kitaev materials.

Finally, we conclude our derivation by noting that a chiral interaction similar to Ref.~\citep{Motrunich2006} will also appear in a minimal model 
including second-neighbor tunneling in the tight-binding Hamiltonian. The hopping on the effective triangular (non-bipartite~\cite{Sen1995}) lattice will lead to a three-spin interactions including $J^{\prime}\sin\left(\phi_{\triangle}\right)\mathbf{s}_{1}\cdot\left(\mathbf{s}_{2}\times\mathbf{s}_{3}\right)$ with $\phi_{\triangle}$ the flux enclosed by the triangle spanned by the three sublattice sites of a honeycomb plaquette. A recent study of minimal model derivation aided by ab inition techniques \citep{Riedl2019} showed that Peierls phases indeed induces three-spin interactions on Kitaev materials. However, they were found to be very small to play a significant role for $\alpha$-RuCl$_3$ and included other interactions besides the chirality term \citep{Riedl2019}. In the following, we will analyse only the effect of the magnetic field over the two-spin interactions.

\begin{center}
\begin{figure}
\begin{centering}
\subfloat{\includegraphics[width=0.47\columnwidth]{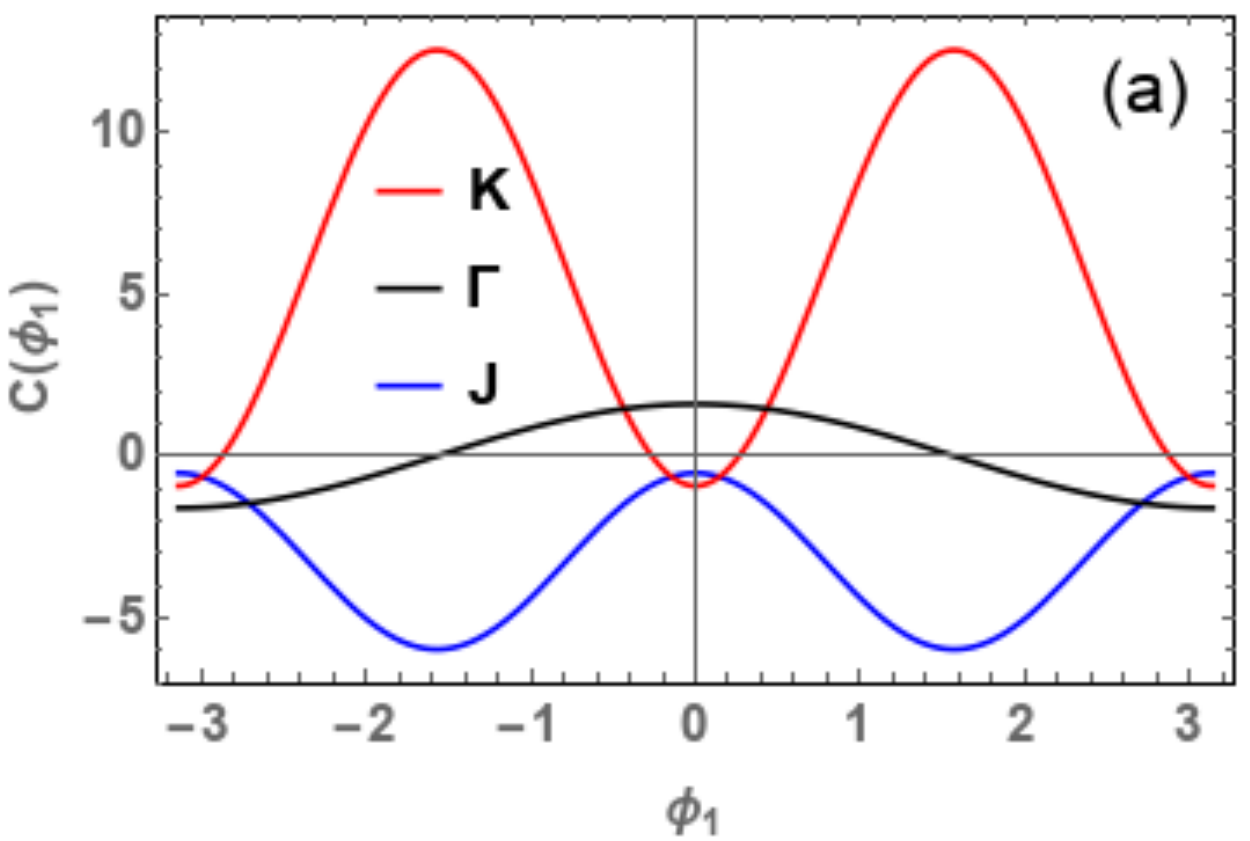}}\subfloat{\includegraphics[width=0.47\columnwidth]{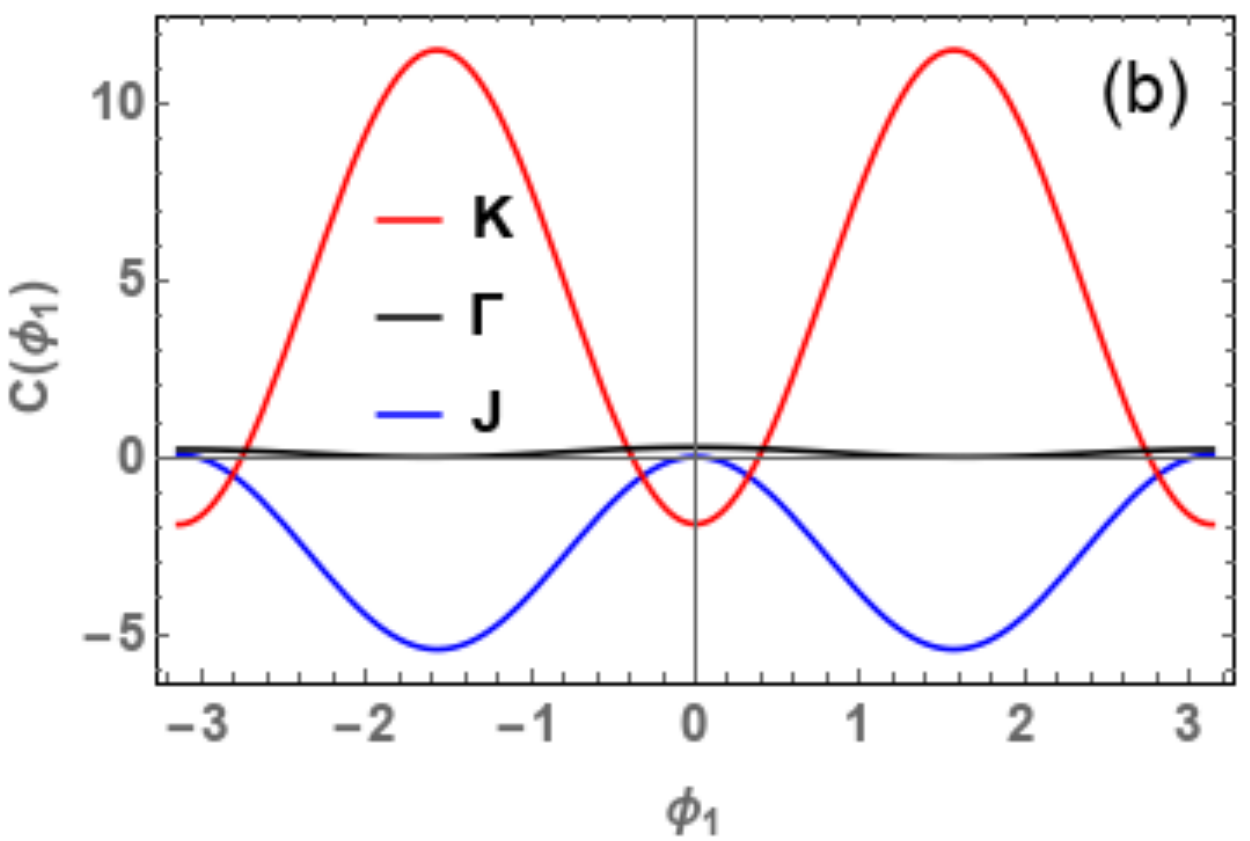}}
\par\end{centering}
\begin{centering}
\subfloat{\includegraphics[width=0.47\columnwidth]{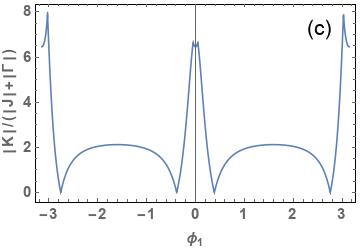}}\subfloat{\includegraphics[width=0.47\columnwidth]{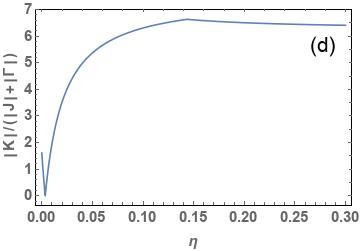}}
\par\end{centering}
\caption{\label{fig:JKG_couplings} (Color online) Oscillatory behavior of
the exchange constants $J\left(\phi_{1}\right)$, $K\left(\phi_{1}\right)$
and $\Gamma\left(\phi_{1}\right)$ with $\eta=0.15$ using (a) the
electronic parameters given by Refs. \citep{Winter2016,Winter2017}
for $\alpha$-RuCl$_{3}$ and (b) the electronic parameters for MOFs
given by Ref. \citep{YamadaPRL2017}. Figures (c) and (d) shows the
value of $\left|K\right|/\left(\left|J\right|+\left|\Gamma\right|\right)$
in MOFs for a fixed value $\eta=0.15$ and a fixed value $\phi_{1}=0.05$,
respectively.}
\end{figure}
\par\end{center}

\emph{Field Modulation of Exchanges.} 
The change of the exchange
constants in Eq. (\ref{JKG_constants_phi}) as a function of the Peierls
phase $\phi_{1}$ and interaction ratio $\eta$ is summarised in Fig. \ref{fig:JKG_couplings}.
In Fig. \ref{fig:JKG_couplings}(a) we use $\eta=0.15$ and electronic parameters ($x_{1}\sim5/16$, $x_{3}\sim-15/16$) calculated with \emph{ab
initio} techniques for $\alpha$-RuCl$_{3}$ as given by Refs. \citep{Winter2016,Winter2017}. 
A simple figure of merit for how close the model is to the pure Kitaev is the  ratio 
$\left|K\right|/\left(\left|J\right|+\left|\Gamma\right|\right)$, in which $K$, $J$ and $\Gamma$ are given by Eq. (\ref{JKG_constants_phi}). For parameters relevant for $\alpha$-RuCl$_{3}$, this ratio is maximal when $\phi_{1}=\pi/2$
and achieves $\approx2.1$. 

Fig. \ref{fig:JKG_couplings}(b) provides the analysis of exchange
constants with the parameters estimated in Ref. \citep{YamadaPRL2017}
($x_{1}=0.023$, $x_{3}=-0.196$) for MOFs . As indicated by Fig. \ref{fig:JKG_couplings}(c),
the model Hamiltonian for MOFs is closer to the KHM at $\phi_{1}=0$
with $\left|K\right|/\left(\left|J\right|+\left|\Gamma\right|\right)\approx6.5$.
The behaviour of the exchange couplings in the neighbourhood of this
point is discussed in the next section. Fig. \ref{fig:JKG_couplings}(d)
shows that the maximum proximity to the KHM is achieved at $\eta\approx0.15$
for a physically achievable case $\phi_{1}=0.05$ as discussed in the next section, which is consistent
with the Hund's coupling estimation for ruthenium \citep{Winter2016}.

\emph{Experimental Implications.} 
Let us now discuss the relevance
of our results for Kitaev candidate materials. We start
with the experiments that indicate a spin-liquid behaviour of $\alpha$-RuCl$_{3}$
under a strong magnetic field \citep{Sears2017,Baek2017,Hentrich2018,Banerjee2016,Banerjee2017,Banerjee2018}.
Taking the Ru-Cl distance to be approximately $2.4\text{Å}$ \citep{KimPRB2016}
and $B\sim$10T, Eq. (\ref{eq:phi}) gives a phase difference of $\phi_{1}-\phi_{2}\sim10^{-2}-10^{-3}$.
Using once again the \emph{ab initio} parameters provided by Refs.
\citep{Winter2016,Winter2017} and setting $\eta=0.15$, the exchange constants  
remain practically constant for this magnetic field: $\left[C\left(\phi_{1}\right)-C(0)\right]/C\left(0\right)\sim10^{-4}-10^{-6}$, 
in which $C$ is $K$, $J$ or $\Gamma$ in Eq. (\ref{JKG_constants_phi}).
For this field strength the energy scale of the diamagnetic term $h\left(\phi_{1}\right)$
is $\approx 10^{-2}-10^{-3}$meV, which is much lower than the scale of the Zeeman
effect ($\sim1\text{meV}$) and the exchange couplings ($\sim1-10\text{meV}$)
\citep{Winter2016}. Therefore, the effects of the Peierls substititution
are too small to play a relevant role in the phase transition observed
in $\alpha$-RuCl$_{3}$. However, for studies of the high-field behaviour our
mechanism becomes relevant for this material. Eq. (\ref{JKG_constants_phi})
indicates that $J\left(\phi_{1}\right)$ will be reduced by 4\% under
a magnetic field of $B\approx110$T while $K\left(\phi_{1}\right)$
and $\Gamma\left(\phi_{1}\right)$ remain practically constant. Although
it is not feasible to implement a static magnetic field with this
intensity, these values are achievable in pulsed high-magnetic field
facilities~\cite{opherden2019magnetization} and will be relevant for studying the high field polarised phase, for example for accurately extracting the values of the exchange constants.

\begin{figure}
\begin{centering}
\includegraphics[width=0.5\columnwidth]{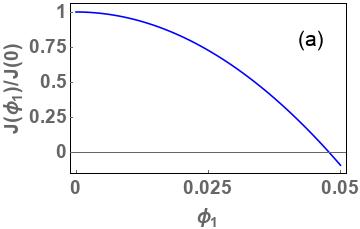}\includegraphics[width=0.5\columnwidth]{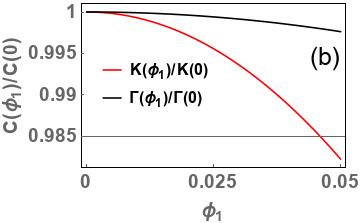}
\par\end{centering}
\caption{\label{fig:couplingvsphi} (Color online) Ratios $C(\phi_{1})/C(0)$
for Kitaev materials based in metal-organic frameworks for (a) the
Heisenberg coupling and (b) the Kitaev and symmetric exchange couplings.}
\end{figure}

The larger area $\mathcal{A}$ of MOF systems makes them a much better
platform to observe the modulation mechanism. Fig. \ref{fig:couplingvsphi}
displays the ratios $C\left(\phi_{1}\right)/C\left(0\right)$ of the
exchange constants with and without an applied magnetic field. The
Heisenberg exchange constant in MOFs also decays much faster than
the Kitaev or symmetric exchange couplings, vanishing at $\phi_{1}\approx0.0479$.
This phase can be achieved by applying a (pulsed) magnetic field $B\sim100$T
over an area $\mathcal{A}\sim31.5\text{Å}^{2}$, which is fairly close
to the values of $\mathcal{A}$ expected in MOFs \citep{YamadaPRL2017,YamadaPRB2017}.
Using $\mathcal{A}\sim31.5\text{Å}^{2}$, we also predict a reduction
of the Heisenberg coupling by 10\% under an achievable static magnetic
field $B\sim30$T. This reduction could drive an observable phase
transition from an ordered phase to a classical spin liquid \citep{Rousochatzakis2017}.

Although the phase $\phi_{1}$ is restricted to small values in Kitaev
materials, we expect that a much broader range can be controlled in
cold-atom implementations of the Kitaev model with similar effects
on the exchange couplings. This prediction is based upon results for
the minimum spin model in optical lattices with artificial spin-orbit
coupling. The effect of this spin-orbit coupling was added in terms
of a Peierls substitution that led to a quantum compass model with
modulated exchange couplings such as in our present study \citep{Radi2012}.
Similar effects could be observed on KHM realizations in cold atoms
\citep{Duan2003,Micheli2006} designed in a way that allows the inclusion
of a Peierls phase. For example, the analogous of $\phi_{1}$ in these
systems can be controlled with appropriate laser setups \citep{Campbell2011,Radi2012}
or using rotating optical lattices \citep{Lewenstein2007,Cooper2008}.
The minimal model in these systems would display exchange constants
that vary similarly to Figs. \ref{fig:JKG_couplings}(a) and (b),
hence providing tunable experimental settings to study quantum phase transitions.

\emph{Conclusion.} We have studied the effect of an applied magnetic field on the low energy description of Mott insulators in the strong SO coupling limit.
We have concentrated our microscopic calculations on Kitaev materials and uncovered a generic mechanism to modulate effective exchange constant through the application of a strong magnetic field.
Since our results are ultimately based only on the existence of more than one exchange
path connecting the interacting sites, this mechanism would be at play in
all instances of  Kitaev materials, e.g. also in three-dimensional extensions
of the honeycomb lattice \citep{Modic2014,Takayama2015}. Similarly, the magnetic
field effects on the hopping integrals would lead to an analogous modulation
of exchange constants in Kitaev materials based on $f$ \citep{Jang2018,Luo2019}
or $d^{7}$ magnetic species \citep{Liu2018,Sano2018}. We propose
that the exchange coupling modulation could be observed with current experimental setups in pulsed
magnetic field facilities, and possibly with static magnetic fields
for MOF systems. The same physical principles could be engineered
in optical lattice realizations of the Kitaev model \citep{Duan2003,Micheli2006}
and provide a rich platform to study quantum phase transitions.

Finally, it would be desirable to extend our  effective low energy models in a field to other correlated quantum materials with sizeable SO couplings, such as parent compounds of iron based superconductors, or effective superlattice structures in the Mott regime where the flux per plaquette is large already for moderate magnetic fields.  

\emph{Acknowledgements.}
WMHN and JK acknowledge support from the Royal Society via a Newton International Fellowship. RM acknowledges
support from the DFG through SFB 1143 (project-id 247310070) and  \textit{ct.qmat} (EXC 2147, project-id 39085490).

\bibliographystyle{apsrev4-1}
\bibliography{RuCl3}

\end{document}